\documentclass[prb,twocolumn,showpacs,groupedaddress,floatfix]{revtex4}
\usepackage{epsfig}
\usepackage{amsmath}
\usepackage{amsfonts}
\usepackage{amssymb}
\usepackage{graphicx}
\begin{document}
\clubpenalty5000\widowpenalty5000
\title{Crossover from Poisson to Wigner-Dyson Level Statistics in Spin Chains with
Integrability Breaking\footnote{Submitted to {\it Physical Review} B}}
\author{D.A. Rabson}
\affiliation{Department of Physics, University of South
Florida, \\ Tampa, FL 33620, USA} 
\author{B.N. Narozhny}
\affiliation{Condensed Matter Section, ICTP, \\ Strada Costiera 11,
I-34100, Trieste, Italy} 
\author{A.J. Millis} 
\affiliation{Physics
Department, Columbia University, \\ 538 West 120th Street, New York,
NY 10027, USA}

\begin{abstract}
We study numerically the evolution of energy-level statistics as an
integrability-breaking term is added to the XXZ Hamiltonian. For
finite-length chains, physical properties exhibit a cross-over from
behavior resulting from the Poisson level statistics characteristic of
integrable models to behavior corresponding to the Wigner-Dyson
statistics characteristic of the random-matrix theory used to describe
chaotic systems. Different measures of the level statistics are
observed to follow different crossover patterns.  The range of
numerically accessible system sizes is too small to establish with
certainty the scaling with system size, but the evidence suggests that
in a thermodynamically large system an infinitesimal integrability
breaking would lead to Wigner-Dyson behavior.

\end{abstract}
\pacs{75.10.Pq, 75.40.Gb}

\maketitle

\section{Introduction}

The conjecture that the statistical properties of energy levels of
chaotic quantum systems may be described in terms of the theory of
random matrices is widely accepted in various fields of
physics. \cite{meh} This however is not a universal property of all
complex interacting systems. One example to the contrary is provided
by the class of the so-called integrable models, \cite{bax} where the
behavior of the system is completely described by a large (infinite in
the thermodynamic limit) set of conserved quantities.  One consequence
is that the level-spacing distribution $P_{\Delta}(E)$ in the case of
integrable models is the Poisson distribution ($\Delta$ denotes mean
level spacing),

\begin{equation}
P_{\Delta}(E)=\frac{1}{\Delta} e^{-E/\Delta} ,\label{Poisson}
\end{equation}

\noindent whereas in random-matrix theory the distribution takes the
Wigner-Dyson form,

\begin{equation}
P_{\Delta}(E) = b_{\beta}\left( \frac{E}{\Delta}\right) ^{\beta}e^{-a_{\beta
}E^{2}/\Delta^{2}} ,\label{Wigner}
\end{equation}

\noindent where $\beta= 1, 2, 4$ correspond to orthogonal, unitary and
symplectic ensembles respectively and \cite{guh} $b_{1}=\pi/2$, $a_{1} =
\pi/4$; $b_{2}=32/\pi^{2}\approx3.24$, $a_{2} = 4/\pi$; $b_{4} =
262144/729\pi^{3} \approx11.6$, $a_{4} = 64/9\pi\approx2.26 $.

Other statistical properties (for example, the evolution of levels
under an external perturbation \cite{faa}) also differ for the two
cases.  One important class of external perturbations is the
application of a voltage.  The difference in response in this case
leads to spectacular differences in transport properties of integrable
and non-integrable models. Integrable models have been argued to have
an infinite conductivity even at high temperatures, essentially
because a typical level has a large response to a voltage, whereas
non-integrable models have a finite conductivity because a typical
level has a small response. \cite{zot,zo2}

While these basic properties have been established for the two generic
cases of integrable and non-integrable models, the crossover between
these two limits as an integrability-breaking interaction is turned on
has not to our knowledge been carefully studied, nor have the
implications of the crossover for the finite-size conductivity of
nearly integrable systems been determined.  Two of us, with N. Andrei,
presented a few numerical results in a paper mainly concerned with the
charge transport of integrable systems. \cite{us1} However, the
significance and interpretation of these results was not clear. Song
and Shepelyansky \cite{son} studied the effects of a random potential
on level statistics of 2D interacting Fermions and found evidence for
a localization-delocalization transition. However, in their case, the
physics of the transition is due to the disorder and thus is different
from the situation in integrable models. Earlier work by DiStasio and
Zotos \cite{distasio95} noted a crossover between Poisson to
Wigner-Dyson only for the low-energy part of the spectrum and did not
address scaling with system size.

In this paper we will fill these gaps by providing numerical results
for finite-size chains with Hamiltonian given by the (integrable) XXZ
model plus an integrability-breaking perturbation $\delta H$. Our
principal results are computations, for finite-length chains, of the
crossover from behavior characteristic of Poisson to behavior
characteristic of Wigner-Dyson statistics in various statistical
measures.  These crossovers do not display an obvious universality in
the sense that different measures show different behavior depending on
the XXZ asymmetry parameter and system size.

Our computations are performed for finite-size systems.  An important
issue is the behavior in the limit of thermodynamically large
system. Extrapolation to the thermodynamic limit proves to be
ambiguous for most of the measures we employ (namely, we cannot rule
out a saturation of the crossover scales as functions of the system
size for chains much longer than those considered in this study), but
the data suggest that all the crossover scales vanish at infinite
system size.

The rest of the paper is organized as follows. First we discuss the
model used in numerical calculations and in particular define
numerically the value of the integrability-breaking parameter at which
a gap appears in the spectrum.  All further considerations will be
devoted to the gapless regime. Then we discuss the level-spacing
distribution and the correlator of level velocities.  The latter is
related to the parametric statistics of the system and also to its
transport properties.  For disordered systems, the correlator of level
velocities was shown to correspond to the dimensionless conductance of
the system, while if one restricts the analysis to periodic boundary
conditions only (see below) it coincides with the Drude
weight. Discussion of the Drude weight concludes the paper.

\section{The model}

We study the effect of integrability breaking on the physical
properties of a spin chain. The integrable model we consider is the
XXZ chain defined on a $N$-site ring with periodic boundary conditions
in the presence of external flux $\phi$ threading the ring:

\begin{align}
& H_{XXZ}=\frac{1}{2}\sum\limits_{i=1}^{N} \left(  e^{i\varphi/N} S_{i}^{+}
S_{i+1}^{-} + e^{-i\varphi/N} S_{i}^{-} S_{i+1}^{+} \right) \nonumber\\
& \nonumber\\
& \quad\quad\quad\quad\quad+ \sum\limits_{i=1}^{N} J_{1}S^{z}_{i}S^{z}
_{i+1}.\label{xxz}
\end{align}

\noindent(Alternatively, the flux can be gauged out to the boundary,
resulting in twisted boundary conditions.) As is well-known,
statistical properties of integrable models are governed by the
Poisson distribution, Eq.(\ref{Poisson}).  Transport properties of the
model can also be inferred from studying the energy levels of the
model, namely by their response to the flux $\phi$. At zero
temperature the behavior of the ground-state energy of the system
under slow variations of the flux determines the Drude weight or the
stiffness \cite{kon} $\mathcal{D}_{s}$ as

\[
\mathcal{D}_{s} = \left.  \frac{N}{2}\frac{\partial^{2}E_{0}}{\partial\phi
^{2}} \right| _{\phi\rightarrow0}.
\]

\noindent Non-vanishing $\mathcal{D}_{s}$ signals ballistic transport
in the system. For the XXZ model at $T=0$ this is the case \cite{sha}
for $-1<J_{1}<1$, where excitations of the system are gapless. If
$|J_{1}|>1$, then the excitation spectrum of the model is
gapped,\cite{descloiseaux} and $\mathcal{D}_{s}=0$. At finite
temperatures the above expression for the stiffness can be generalized
\cite{zot,zo2,us1} to $\mathcal{D}_{s}=D_{1}+D_{2}$, where

\[
D_{1} = - \frac{N}{2\beta} \frac{1}{\mathcal{Z}}\left.  \frac{\partial^{2}
\mathcal{Z}}{\partial\phi^{2}}\right| _{\phi\rightarrow0}
\]

\noindent vanishes in the thermodynamic limit, \cite{gin} and the rest is
\textit{positive}:

\begin{equation}
D_{2} = \frac{\beta N}{2} \frac{1}{\mathcal{Z}} \sum\limits_{n} \left.  \left(
\frac{\partial E_{n}}{\partial\phi}\right) ^{2} \right| _{\phi\rightarrow
0}\displaystyle{e^{-\beta E_{n}}}.\label{D2}
\end{equation}

\noindent In the gapless phase of the XXZ model it has been shown
\cite{zot,zo2,us1} that ballistic transport persists to finite
temperatures in the sense that
$D(N)=\lim_{T\rightarrow\infty}D_{2}(N)T>0$. The infinite-temperature
limit of this result implies that for a typical level
$dE_{n}/d\phi\sim1/\sqrt{N}$. At the antiferromagnetic Heisenberg
point $J_{1}=1$ the model still has gapless excitations, but results
of Fabricius and McCoy \cite{mac} suggest that $D_{2}$ vanishes
(slowly) as the system size increases. Numerical results of Narozhny
\textit{et al.}\cite{us1} were consistent with this suggestion, but
the limited range of system sizes attainable precluded a definite
statement.

Integrability breaking is introduced by adding the term with
next-neighbor coupling

\begin{equation}
\delta H = \sum\limits_{i=1}^{N} J_{2}S^{z}_{i}S^{z}_{i+2} \quad.\label{j2}
\end{equation}

\noindent This term should be contrasted to that considered by Eggert
\cite{egg} insofar as it is explicitly not SU(2) invariant. However,
away from the Heisenberg point the effect of the interaction
Eq.~(\ref{j2}) is similar to that of its SU(2)-invariant counterpart:
(i) it breaks the integrability of the system, and (ii) for large
enough values of $J_{2}$ it causes the system to dimerize, so that the
spectrum acquires a gap. The critical value of $J_{2}$ at which the
gap opens is of course different from the $0.24$ found in
Ref.~\onlinecite{egg}. Our numerical estimates \cite{est} suggest a
value $J_{2}^{(c)}\gtrsim1.1$ that is a weakly increasing function of
both system size and $J_{1}$: for $N=18$ and $J_{1}=0.2$, for example,
the gap appears to open at $J_{2}=1.13\pm0.01$. The gap opening limits
the range of values of $J_{2}$ under consideration, as we are
interested only in properties of the gapless phase; indeed, the
measures we consider presently begin showing different behavior for
$J_{2}>J_{2}^{(c)}$. Similarly, the parameter region considered by
Faas \textit{et al.}\cite{faa} belongs to the gapped regime, which
accounts for certain differences in the behavior of the level
statistics reported in Ref.~\onlinecite{faa} and in the present paper.

In this paper we study the eigenvalues of $H=H_{XXZ}+\delta H$ and
their evolution under the change of $\phi$ for the above models with
varying $J_{1,2}$ and system size. For the non-integrable Hamiltonian
$H_{XXZ}+\delta H$ we use exact numerical diagonalization to construct
the level-spacing distribution and level auto-correlation functions
and to evaluate the stiffness $D_{2}$, Eq.~(\ref{D2}). The use of
exact numerical methods is motivated by the need to obtain the whole
spectrum of the model in order (i) to study the statistical properties
of the spectrum and (ii) to study the stiffness Eq.~(\ref{D2}) at
infinite temperature. The drawback of the method is the limitation to
small system sizes (we present results for chains of up to $20$
sites).  For finite system sizes we obtain a detailed characterization
of the crossover.

{ \begin{figure}[th]
\epsfxsize=7 cm \centerline{\epsfbox{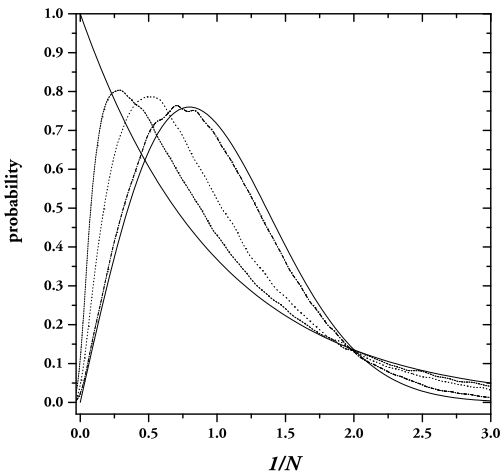}}
\caption{Typical crossover of the level-spacing distribution from
Poisson (left solid curve) for $J_{2}=0$ to Wigner-Dyson (right solid
curve) for a representative system. The plot is made for $N=18$,
$S^{z}=3$, $J_{1}=0.2$, and $J_{2}=0.1, 0.2, 0.5$. For $J_{2}=0$ the
numerical distribution agrees very closely with the exponential
plotted. The Wigner-Dyson distribution shown is the theoretical curve
for the orthogonal ensemble.}%
\label{pe}%
\end{figure}}

\section{Level-spacing distribution}

We begin with a brief discussion of the integrable case. The
level-spacing distribution for $J_{2}=0$ is the Poisson distribution
(shown in Fig.\ref{pe} by the left solid curve). This illustrates the
fact that the integrable system has so many conservation laws that
levels essentially do not repel each other.  To characterize transport
properties of the system we show in Fig.~\ref{figd2} the quantity
$D(N)=D_{2}(N)T$ at $T\rightarrow\infty$ for different system sizes
and different values of the integrable interaction $J_{1}$ (dashed
lines in Fig.~\ref{figd2}). $D$ is seen to be almost size-independent
for the cases $J_{1}<1$, in agreement with previous work, \cite{us1}
while a weak size dependence is evident in the Heisenberg case
$J_{1}=1$. Although this dependence appears to have a positive
$y$-intercept, we believe that the system size in this study is still
too small to make a definite statement regarding the behavior of the
Heisenberg model in the thermodynamic limit.\cite{twothirds}

{ \begin{figure}[th]
\epsfxsize=7 cm \centerline{\epsfbox{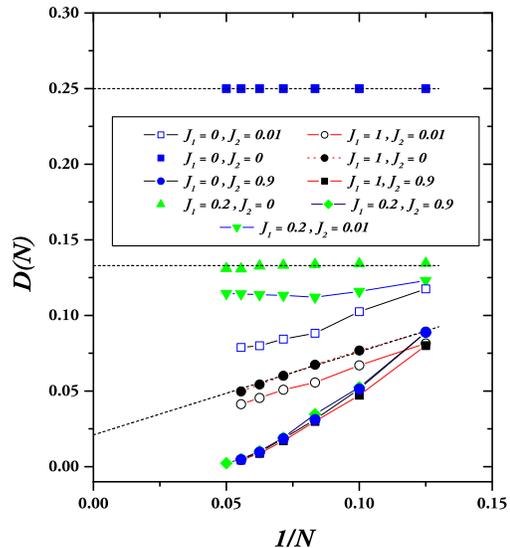}}
\caption{The stiffness $D(N)=\lim_{T \rightarrow \infty} D_2(N)T$ as a
function of inverse system size for different values of interaction
parameters. Dashed lines show a naive extrapolation to the
thermodynamic limit for the integrable system. Here we show the
results for $J_{1}=0$ (squares), $J_{1}=0.2$ (triangles) and $J_{1}=1$
(circles). Empty symbols correspond to the smallest $J_{2}=0.01$,
which seems to affect strongly only the $J_{1}=0$ case. The behavior
for $J_{2}=0.9$ appears to be independent of $J_{1}$. }%
\label{figd2}%
\end{figure}}

We turn now to the case of broken integrability. As the
integrability-breaking term Eq.~(\ref{j2}) is added to the
Hamiltonian, energy levels immediately start to repel,\cite{rep} and
as a consequence immediately $P_{\Delta}(0)=0$ so that the
distribution acquires a peak. As illustrated in Fig.~\ref{pe},
increasing $J_{2}$ shifts the peak to the right until the distribution
starts to look like the Wigner-Dyson distribution (shown in
Fig.~\ref{pe} by the right solid curve). At the same time the tail of
the distribution changes from the exponential in Eq.~(\ref{Poisson})
to the (asymptotically) Gaussian tail of the Wigner-Dyson
distribution.

To quantify this crossover we show the evolution of the peak position
and the characteristics of the tail with the change in $J_{2}$ in
Figs.~\ref{peak} and \ref{tail}. Both exhibit similar features,
although the estimates for the crossover scales extracted from the two
are numerically different (see table~\ref{jtab} and insets in
Figs.~\ref{peak} and \ref{cumul}).

{ \begin{figure}[th]
\epsfxsize=7 cm \centerline{\epsfbox{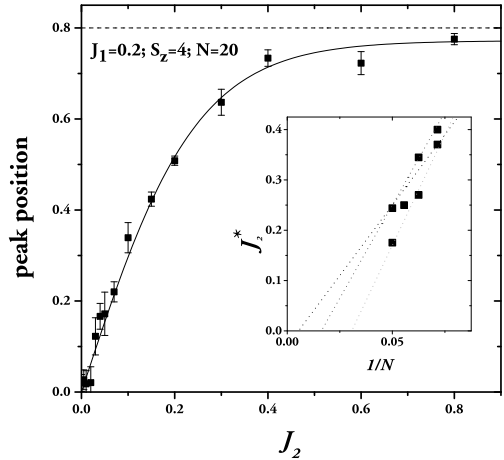}}
\caption{Typical crossover of the peak position. The data correspond
to $N=20$, $S^{z}=4$, $J_{1}=0.2$, with the solid line a fit to the
form $a\tanh(J_2/J_2^*)$, $J_2\approx0.25$.  The dashed line indicates
the peak position of the Wigner-Dyson distribution. The inset shows
finite-size scaling of the crossover scale (the data points correspond
to $N=20$, $S^{z}=4$; $N=20$, $S^{z}=3$; $N=18$, $S^{z}=3$; $N=16$,
$S^{z}=3$; $N=16$, $S^{z}=2$; $N=14$, $S^{z}=3$; $N=14$, $S^{z}=2$);
the straight lines are guides to the eye, suggesting that $J_{2}^{*}$
vanishes for the infinite chain.}%
\label{peak}%
\end{figure}}

As shown in Fig.~\ref{peak}, the peak of the distribution grows from
zero to the value characteristic of the Wigner-Dyson distribution and
then saturates.  To estimate the crossover scale $J_{2}^{*}$, we fit
the data by the hyperbolic tangent \cite{fit} of the form $a
\tanh(x/x_{0})$ with $x_{0}$ approximating $J_{2}^{*}$. The inset
shows the resulting values for $J_{2}^{*}$ as a function of the system
size. As we noted before, \cite{rep} we are restricting our attention
to fixed values of the total spin $S^{z}$. However, for the purposes
of the finite-size scaling, it makes more sense to compare data with
the fixed ratio $S^{z}/N$. One way to see this is to recall that by
means of the Jordan-Wigner transformation the spin chain can be mapped
onto a system of spinless Fermions.\cite{us1} In the Fermion language,
$1/2-S^{z}/N$ corresponds to the filling fraction. Since it is not
possible to keep the ratio $S^{z}/N$ exactly the same for all values
of $N$ used in this paper, we choose to present the data for two
sectors of fixed $S^{z}$ that are closest to the chosen value of
$S^{z}/N$. Therefore the inset in Fig.~\ref{peak} shows two data
points for the $N$ other than $N=18$ (we chose $S^{z}/N=1/6$). The
straight lines are just guides to the eye.

To analyze the evolution of the tail, we approximate the intermediate
distributions (see Fig.~\ref{pe}) by

\[
P_{\Delta}(E)\propto\exp\left[ -a \frac{E}{\Delta} - b \left( \frac{E}{\Delta
}\right) ^{2}\right] .
\]

\noindent Clearly, for Eq.~(\ref{Poisson}) $a=1$ and $b=0$, while for
the orthogonal ensemble, Eq.~(\ref{Wigner}) corresponds to $a=0$ and
$b=\pi/4$. In Fig.~\ref{tail} we show the evolution of $b$ (the fact
that plotted values never reach $\pi/4$ is an artifact of the
calculation). Fitting the curve to a hyperbolic tangent, we can
extract an estimate for the crossover value
$J_{2}^{\ast}(N=20)=0.27$. This value differs somewhat from the one
extracted from the peak position (for the same values of $N$, $S^{z}$,
and $J_{1}$); see the inset in Fig.~\ref{peak}. The behavior of the
tail characteristics with respect to changing system size exhibits the
same trend as shown in the inset in Fig.~\ref{peak} for the peak
position: the characteristic scales tend to decay with increasing
system size. The naive extrapolation of such a trend is consistent
with a statement of vanishing $J_{2}^{\ast}$ as $N\rightarrow \infty$;
however, the data are insufficient to prove it.

{ \begin{figure}[th]
\epsfxsize=7 cm \centerline{\epsfbox{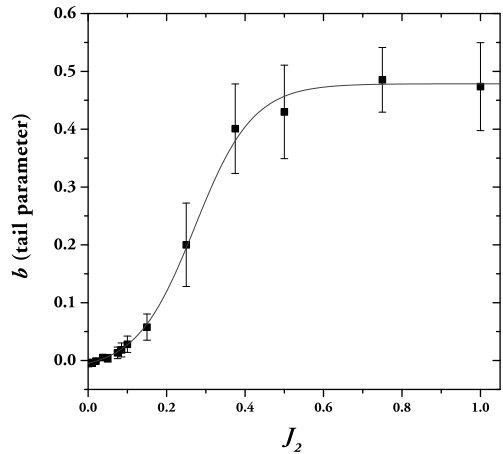}}
\caption{Evolution of the parameter $b$ from the tail of the
level-spacing distribution. For the integrable case, $b=0$. For large
$J_{2}$ it does not quite reach the Wigner-Dyson value $\pi/4$, but it
still shows a clear crossover. The crossover scale estimated by
fitting the data to a hyperbolic tangent is ${J_{2}^{*}}=0.27$. The
plot is made for $N=20$, $S^{z}=3$, $J_{1}=0.2$.}%
\label{tail}%
\end{figure}}

Another way to quantify the evolution of the level-spacing
distribution shown in Fig.~\ref{pe} is to consider
cumulants. \cite{cum} Their unbiased estimators (the Fisher statistics
\cite{fis} $k_{n}$) are easily computed. For our normalized level
spacings, the first cumulant (which is equal to the mean) is unity.

The cumulants of a distribution characterize its width (second
cumulant, or variance) and shape;\cite{gau} beyond perhaps the fifth,
numerical cumulants become too sensitive to outliers to be of much
use. A study of the cumulants of a distribution is qualitatively
similar to our foregoing study of the tails, but it turns out to be
simpler numerically. In Fig.~\ref{cumul} we show the unbiased variance
estimate, $k_{2}$, as a function of $J_{2}$ for system size $N=20$,
$J_{1}=0.2$, $S^{z}=3$, momentum $k=0$. The theoretical limits should
be $1$ for the Poisson distribution and $\pi/4-1=0.2732$ (the bottom
of the scale) for Wigner-Dyson. The fact that the data points deviate
from these ideal values is an artifact of finite sampling. Fitting the
curve to a hyperbolic tangent, we estimate the crossover scale, shown
in the inset as a function of size (for the same sequence of quantum
numbers as in Fig.~\ref{peak}). The inset also shows the crossover
scale for the fourth cumulant. As before, the scaling suggests (but
does not establish) that the crossover scale $J_{2}^{*}$ associated
with either cumulant should vanish in the limit of infinite size. It
is not possible to determine whether the different measures, $k_{2}$
and $k_{4}$ (we also looked at $k_{3}$), scale in the same way or
differently with system size.

{ \begin{figure}[th]
\epsfxsize=7 cm \centerline{\kern-2em\epsfbox{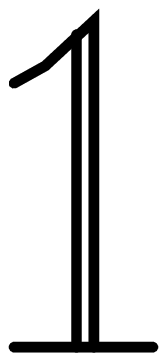}}
\caption{Variance estimate ($\times$), $k_{2}$, as a function of
$J_{2}$ for system size $N=20$. ($J_{1}=0.2$, $S^{z}=3$.) The
crossover scale is estimated as the turning point ($\circ$) in a tanh
fit (curve). The inset illustrates the finite-size scaling for
cumulants $k_{2}$ (solid lines and $\times$) and $k_{4}$ (dotted lines
and $\diamond$) for the same set of sizes and $S^{z}$ sectors as in
Figs.~\ref{peak} and \ref{c0} appropriate for 1/3 filling. The
$J_{2}^{*}$ associated with either cumulant may vanish in the limit of
infinite size.}%
\label{cumul}%
\end{figure}}

The second cumulant $k_{2}$ plays a role similar to that of the
parameter $\eta$ used in Ref.~\onlinecite{son} to estimate the overall
``proximity'' of the observed distribution to either the Poisson or
the Wigner-Dyson limit.  In that sense, Fig.~\ref{cumul} shows
behavior similar to that found in Ref.~\onlinecite{son}, although the
physics of the evolution of levels is quite different in our case.

\section{Elements of the parametric statistics}

More information about the crossover to the chaotic behavior described
by the Wigner-Dyson statistics can be extracted from the study of the
autocorrelation functions. Here we will discuss the autocorrelation of
level velocities

\begin{equation}
C(\phi) = \frac{1}{\Delta^{2}} \left\langle \frac{\partial E_{i}(\theta
)}{\partial\theta} \frac{\partial E_{i}(\theta+ \phi)}{\partial\theta}
\right\rangle _{\theta,i},\label{c}
\end{equation}

\noindent where the angular brackets indicate averaging over a set of levels
and fluxes.\cite{faa}

{ \begin{figure}[th]
\epsfxsize=7 cm \centerline{\kern-1.5em\epsfbox{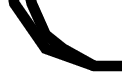}}
\caption{Autocorrelation of level velocities for various values of
integrability breaking. Only a short range of flux is shown; in all
cases, the sum rule (the integral of $C(\phi)$ over all fluxes
vanishing) is satisfied numerically. For $J_{2}>0$, there is a point
of inflection at non-zero flux.  However, the integrable model has no
initial inflection.}%
\label{cphi}%
\end{figure}}

A typical form of the autocorrelation function Eq.~(\ref{c}) is shown
in Fig.~\ref{cphi}. For large values of $J_{2}$, this form resembles
the universal correlator characteristic of chaotic systems. \cite{faa}
However, in the crossover region, $C(\phi)$ deviates from the
universal form in a rather complex fashion, which makes a quantitative
analysis of the crossover difficult. Therefore we focus on two
particular features of the curve, the turning point and $C(0)$. We
note a feature of the autocorrelation curve shown in Fig.~\ref{cphi}
that appears only as the integrability breaking is introduced: all
curves for $J_{2}>0$ have a non-zero point of inflection as the
autocorrelation decreases with the increase of $J_{2}$, but the
autocorrelation function of the integrable system does not have such
an inflection point. Consequently, the behavior near zero flux changes
from linear ($C(\phi)-C(0) \propto- \phi$) for the integrable case to
quadratic for $J_{2}>0$.

The autocorrelation function of level velocities at zero flux
difference $C(0)$ is somewhat similar to the stiffness Eq.~(\ref{D2}),
the differences being that $C(0)$ is also averaged over a set of
fluxes, does not contain the extra factors of temperature and system
size, and corresponds to a single sector of fixed $S^{z}$. However, in
chaotic systems it is $C(0)$ that can be related to
transport. \cite{faa} There it was argued to correspond to the
dimensionless conductance.

{ \begin{figure}[th]
\epsfxsize=7 cm \centerline{\epsfbox{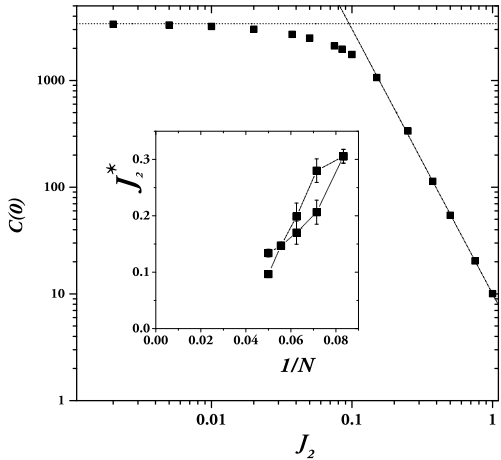}}
\caption{$C(0)$ as a function of $J_{2}$ for $N=20$, $J_{1}=0.2$,
$S^{z}=3$.  The dotted line corresponds to the value at $J_{2}=0$. The
inset shows the finite-size scaling of the crossover scale,
$J_{2}^{*C_{0}}$, at approximately fixed ratio $S^{z}/N=1/6$ (for
$N=14$, $16$, or $20$, we pick the two closest values of $S^{z}$ for
these sizes). }%
\label{c0}%
\end{figure}}

In Fig.~\ref{c0} we show the behavior of $C(0)$ as a function of
$J_{2}$.  Clearly, for finite systems, $C(0)$ exhibits a well-defined
crossover. For the dataset presented in Fig.~\ref{c0} ($N=20$,
$J_{1}=0.2$, $S^{z}=3$), the correlator $C(0)$ decays as a
$\approx5/2$-power law after $J_{2}$ exceeds the value
$J_{2}^{*C_{0}}=0.097\pm0.004$ (defined in Fig.~\ref{c0} as a crossing
point of the above power law --- the straight line in the log-log
scale --- with the value at $J_{2}=0$).

The inset shows the crossover scale as a function of the size (in the
same manner as the crossover scale extracted from the peak of the
level-spacing distribution). The behavior is very similar to that in
the inset in Fig.~\ref{peak} (although numerical values of the
crossover scales differ in the two cases). Both would be consistent
with the statement that $J_{2}^{*}\rightarrow0$ as
$N\rightarrow\infty$; however such a conclusion cannot be ascertained
on the basis of the data available.

\section{Spin stiffness}

Now we discuss the effect of the integrability breaking Eq.~(\ref{j2})
on the stiffness $D_{2}$. In Fig.~\ref{figd2} we show three sets of
data corresponding to three different values of the XXZ anisotropy
parameter $J_{1}$. The data illustrate the following tendencies:

(i) For $0<J_{1}<1$ (represented by $J_{1}=0.2$; similar behavior is
observed for other values) the data clearly show that very small
integrability breaking (characterized by $J_{2}=0.01$) has little
effect on the stiffness of the finite chains (which is to be
expected). Moreover, the extrapolation to infinite size seems to
result in a finite value for the stiffness in a manner similar to that
of the integrable model.

(ii) For the Heisenberg model, small integrability breaking again does
not have a pronounced effect; however, in this case (even though the
extrapolation indicates a finite value for the thermodynamic limit),
one cannot be certain of the behavior of the infinite chain.

(iii) For $J_{1}=0$ the situation is different: even a very small
amount of integrability breaking leads to a sharp reduction in the
stiffness for finite chains. The extrapolation to infinite size is
also uncertain. It should be noted, however, that if one compares the
behavior of two integrable cases, $J_{1}=0$ and $J_{1}>0$, then a
similar picture arises (compare for example the two top dashed lines
in Fig.~\ref{figd2}). This has to do with the fact that the spin chain
at $J_{1}=0$ can be mapped (by means of the Jordan-Wigner
transformation) onto a system of free spinless Fermions and as such
possesses more symmetries than even the integrable (but interacting)
XXZ model.

(iv) When the integrability-breaking parameter is not small, the
stiffness decays sharply (in fact, if we were to show a log-log plot,
faster than any power law) with system size and clearly extrapolates
to zero in the thermodynamic limit. This behavior is qualitatively
independent of $J_{1}$.

{ \begin{figure}[th]
\epsfxsize=8.5 cm \centerline{\epsfbox{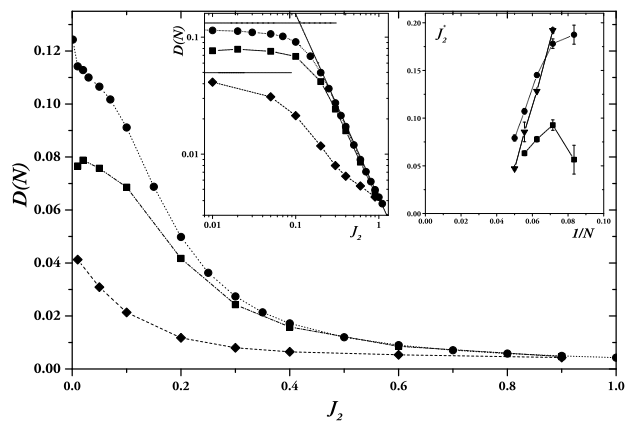}}
\caption{$D(N)$ as a function of $J_{2}$ (for $N=18$).  The main plot
and the left inset show the same data in linear and log-log scale
(meant to illustrate the crossover to a power-law decay of the
stiffness at large $J_{2}$ for smaller values of $J_{1}$) for
$J_{1}=0$ (squares), $J_{1}=0.2$ (circles), and $J_{1}=1$
(diamonds). The horizontal lines in the inset indicate the values of
$D(N)$ at $J_{2}=0$ for $J_{1}=0.2$ (the upper line) and
$J_{1}=1$. The second inset shows the size dependence of the crossover
scale for $J_{1}=0$, $0.2$, and $0.4$ (triangles). The Heisenberg case
$J_{1}=1$ does not show a clear crossover. }%
\label{d2j2}%
\end{figure}}

Thus the data suggest that for finite chains there exists a
``critical'' value of $J_{2}$ (smaller than the point of the gap
opening) beyond which the stiffness tends to vanish. This value is not
universal, in the sense that it depends on $J_{1}$. This situation is
illustrated in Fig.~\ref{d2j2}, where we show the dependence of the
stiffness on $J_{2}$ for three values of $J_{1}$ and the fixed system
size $N=18$. For large values of $J_{2}$, all three curves saturate to
zero (although the one with $J_{1}=1$ does so faster). For $J_{1}=0.2$
the effect of small $J_{2}$ is rather weak, and the curve exhibits a
clear crossover. For $J_{1}=0$ the crossover also appears; however,
the value $0.08$ to which the curve tends as $J_{2}\rightarrow0$ is
much smaller than the value at $J_{2}=0$ exactly (which is $0.25$ and
is thus outside the frame of the plot). The crossover is illustrated
in the left inset. The right inset shows the change of the crossover
scale with system size for $J_{1}=0$, $0.2$, and $0.4$. The behavior
at $J_{1}=1$ is quite different and in particular does not show an
obvious crossover (and is thus not represented in the right
inset). This behavior might be related to the conjecture \cite{mac}
that at the Heisenberg point the stiffness vanishes in the
thermodynamic limit even without the integrability
breaking. Alternatively, this can reflect the fact that the Heisenberg
model is characterized by logarithmic correlations,\cite{gin} and thus
the small systems considered in this paper are not representative.

\section{Discussion}

Prior to performing the calculations one could have had two
conflicting expectations for the behavior of the non-integrable
system: (i) as soon as the integrability is broken the system becomes
chaotic and as a result shows diffusive transport; (ii) there exists a
``critical'' magnitude of the integrability breaking that separates
the chaotic regime from the one that retains some features of the
integrable model, in particular, ballistic transport.

The latter picture has an analogy in the localization problem in
disordered conductors. \cite{alt} The states of an integrable model
can be visualized as well-defined localized points in the
multidimensional space of the integrals of motion characteristic of
the model. These points are well separated due to the quantization of
the values of the integrals of motion. Consider now the effect of an
infinitesimally small integrability breaking. One can certainly expect
the points to spread out into fuzzy spots, but at the same time one
might argue that unless the integrability breaking is strong enough,
these spots do not overlap. In this regime the system retains some
memory of the fact that it was indeed integrable before the extra
interaction was turned on. When the integrability breaking is so
strong that the spots overlap into a continuum, the system becomes
fully chaotic.

The numerical analysis presented in this paper seems roughly
consistent with the second possibility for finite chains: a small
integrability-breaking term leads to behavior that is close to that of
the integrable system.  Quantities related to transport, the stiffness
$D_{2}$ and the \textquotedblleft conductance\textquotedblright,
$C(0)$, exhibit a reasonably rapid crossover as functions of the
strength of the integrability-breaking interaction. The crossover
behaviors seem to be different for different
quantities. Table~\ref{jtab}, for example, illustrates some of this
variability for the example of $J_{1}=0.2$, intermediate between the
non-interacting model and the Heisenberg point.  Similarly, the
quantitative characteristics of the level-spacing distribution (namely
the peak position and the tail parameter, see Figs.~\ref{peak} and
\ref{tail}, or cumulants, see Fig.~\ref{cumul}) exhibit similar
crossovers (the corresponding scales are also included in
table~\ref{jtab}).

The one exception to this picture is the number of degenerate levels
in the system \cite{rep} represented by $P_{\Delta}(0)$. This measure
exhibits a jump as infinitesimally small (numerically meaning of the
order of the computer precision) $J_{2}$ is introduced (namely
$P_{\Delta}(0)$ vanishes, as illustrated in Fig.~\ref{pe}).

\begin{table}[ptb]
\caption{The system crosses over from integrable to fully chaotic
behavior with \textit{different\/} crossover scales, depending on what
is being measured. Furthermore, the crossover scales themselves scale
differently with system size. We calculate crossover scales associated
with peak position, tail crossover (from exponential to Gaussian),
mean squared level velocity $C_{0}$, the fourth cumulant $k_{4}$ of
the level-spacing distribution, and conductance $D_{2}T$. In this
example, $J_{1}=0.2$. The crossover $J_{2}^{*D_{2}T}$ is calculated
for the entire spectrum, while all the others are calculated for the
$S^{z}=3$ sector. Entries of $-$ could not be extracted from the data
because of numerical uncertainty. Rough error estimates for the least
significant digit are provided where available. }%
\label{jtab}%
\begin{ruledtabular}
\begin{tabular}{cccccc}
N & $J_2^{*\rm peak}$ & $J_2^{*\rm tail}$ & $J_2^{*C_0}$
& $J_2^{*k_4}$ & $J_2^{*D_2T}$  \\
\hline
20 & 0.19 & 0.27 & 0.097(4) & 0.091 & 0.079(3) \\
18 & 0.25 & 0.43 & 0.15(1) & 0.17 & 0.107(3) \\
16 & 0.34 & 0.49 & 0.20(2) & 0.24 & 0.145(2) \\
14 & 0.38 & $-$ & 0.28(2) & $-$ & 0.178(5) \\
\end{tabular}
\end{ruledtabular}
\end{table}

Conclusions for the thermodynamic limit are harder to draw from our
data. The variation of crossover scale with system size indicated in
the insets to the different figures suggests that the crossover scale
vanishes in the limit of infinite size system rather than saturating at
non-zero values for $J_2^*$, but the limited range of sizes available
to us, along with the absence of a theoretically justified
extrapolation to the thermodynamic limit, precludes a definite
statement.  Constructing a theory of the approach to the infinite-size
limit of chains with weak integrability breaking remains an important
open problem.

\section*{Acknowledgments}

Instructive discussions with B.L.\ Altshuler, A.A.\ Nersesyan, and
J.K.\ Looper are gratefully acknowledged. Numerical work was performed
at the San Diego Supercomputing Center and the University of Michigan
supercomputing facility through NPACI grant CSD268 and at the
Research-Oriented Computing Center of the University of South
Florida. DAR is a Cottrell Scholar of Research Corporation and also
wishes to thank the Abdus Salam International Centre for Theoretical
Physics for its hospitality.


\begin{thebibliography}{99}
\bibitem {est}The numerical estimates for the opening of the gap are
based on the following observation on the XXZ model, where the gap
opens\cite{descloiseaux} for $|J_{1}|>1$. The ground state of the
system in the gapless regime is unique and corresponds to total spin
$S^{z}=0$. The first excited state is a Heisenberg triplet ($S^{z}=1$)
and is doubly degenerate away from the Heisenberg point. The next
excited state is again a singlet. Now, when the gap opens, the ground
state becomes doubly degenerate in the infinite chain, but it is still
a singlet. Thus, as one varies slowly the interaction parameter
responsible for opening the gap, the non-degenerate excited singlet
has to cross the degenerate triplet in order to merge with the ground
state. For the XXZ model, this happens as $J_{1}$ crosses through the
Heisenberg point, at which the three lowest excited states are
degenerate.  Increasing $J_{1}$ should in the thermodynamic limit lead
to doubly degenerate ground and first excited states. However, for
finite chains this never happens for $J_{1}$ slightly exceeding
$1$. Instead, the spectrum consists of the singlet ground state and
the first excited state with the next excited state being the doubly
degenerate triplet. Increasing $J_{1}$ reduces the interval between
the singlets while increasing the interval between the ``excited''
singlet and the triplet. For fixed $|J_{1}|<1$, the point in $J_{2}$
at which one now observes this crossing is a weakly increasing
function of system size and of $J_{1}$.
\bibitem {twothirds}In computing levels and the other statistics
reported here, we consider only the middle two-thirds of levels,
consistent with Ref.~\onlinecite{faa}. This avoids the
unrepresentative spacings of extremal eigenvalues in finite systems;
qualitative conclusions are independent of the middle fraction ($1/3$,
$2/3$, $3/4$) employed.
\bibitem {rep}Breaking integrability of course does not break the
trivial global conservation laws, such as energy and total spin. In
addition Eq.~(\ref{j2}) preserves the translational invariance of the
XXZ Hamiltonian.  Therefore, in order to observe level repulsion one
has to restrict oneself to a set of levels with particular values of
the total spin $S^{z}$ and momentum.  All the other conservation laws
present in the XXZ model are broken by Eq.~(\ref{j2}). (Parity also is
conserved in the special case $S^{z}=\phi=0$.)
\bibitem {fit}While the choice of the functional forms used to
describe the crossovers is somewhat arbitrary, we choose two
particular ways to estimate the crossover scales. Parameters of the
level-spacing distribution were fitted to a hyperbolic tangent (which
illustrates exponential saturation of the peak position and the tail
form as the distribution approaches the Wigner-Dyson limit), while the
stiffness and level-velocity autocorrelation function were found
numerically to exhibit a crossover to a power-law decay for large
$J_{2}$.
\bibitem {gau}The higher cumulants describe the deviation of a distribution
from a Gaussian, for which they are identically zero.
\end{thebibliography}
\end{document}